\documentclass[a4paper,11pt]{article}
\usepackage{pos}
\usepackage{graphicx}
\usepackage{wrapfig}
\usepackage{multicol}
\usepackage{slashed}
\usepackage{physics}
\usepackage{subcaption}
\usepackage{wrapfig}
\usepackage{multicol}
\usepackage[colorinlistoftodos,prependcaption,textsize=small]{todonotes}
\usepackage[acronym]{glossaries}
\makeglossaries

\newcommand{\Z}{\mathcal{Z}}
\newcommand{\D}{\mathcal{D}}

\newacronym{qcd}{QCD}{Quantum Chromodynamics}
\newacronym{lqcd}{LQCD}{Lattice QCD}
\newacronym{qgp}{QGP}{Quark-Gluon Plasma}
\newacronym{hic}{HIC}{Heavy-Ion Collision}
\newacronym{cep}{CEP}{Critical Endpoint}
\newacronym{eos}{EoS}{Equation of State}
\newacronym{lcp}{LCP}{Line of Constant Physics}

\title{Dense and magnetized QCD from imaginary chemical potential}

\author[a]{S. Bors\'{a}nyi}
\author[b]{B. B. Brandt}
\author[b,d]{G. Endr\H{o}di}
\author[a]{J. N. Guenther}
\author*[a]{M. A. Petri}
\author*[b,c]{A. D. M. Valois}
\author[a]{L. Varnhorst}

\affiliation[a]{Universität Wuppertal, Department of Physics, Wuppertal D-42119, Germany}
\affiliation[b]{Universität Bielefeld, Universitätsstraße 25, 33615 Bielefeld, Germany}
\affiliation[c]{CAFPE and Departamento de Física Teórica y del Cosmos, Universidad de Granada, E-18071
Granada, Spain}
\affiliation[d]{Institute of Physics and Astronomy,
ELTE E\"otv\"os Lor\'and University,\\
P\'azm\'any P.\ s\'et\'any 1/A, H-1117 Budapest, Hungary}

\emailAdd{borsanyi@uni-wuppertal.de}
\emailAdd{brandt@physik.uni-bielefeld.de}
\emailAdd{endrodi@physik.uni-bielefeld.de}
\emailAdd{jguenther@uni-wuppertal.de}
\emailAdd{petri@uni-wuppertal.de}
\emailAdd{dvalois@physik.uni-bielefeld.de}
\emailAdd{varnhorst@uni-wuppertal.de}

\abstract{In this work, we computed the equation of state of dense QCD in the presence of background magnetic fields using lattice QCD simulations at imaginary baryon chemical potential. Our simulations include 2+1+1 flavors of stout-smeared staggered fermions with masses at the physical point and a tree-level Symanzik-improved gauge action. Using several expansion schemes, we tuned our simulation parameters such that the equation of state satisfies strangeness neutrality and isospin asymmetry constraints, which are relevant to the phenomenology of heavy-ion collisions. Our results suggest a strong change in the equation of state due to the magnetic field, in particular, around the crossover temperature. A continuum extrapolation of our data is still needed for future applications of our equation of state to heavy-ion-collision phenomenology.}
\FullConference{The 41st International Symposium on Lattice Field Theory (LATTICE2024)\\
 28 July - 3 August 2024\\
Liverpool, UK\\}

\begin{document}
\maketitle

\section{Introduction}\label{sec:intro}
    
It is widely believed that strongly interacting matter existed as a \gls{qgp} during the first microseconds of the Universe, where quarks and gluons were subjected to high temperatures and densities. During that epoch, the so-called electroweak phase transition may have generated magnetic fields comparable with the typical energy scale of the strong interactions~\cite{Vachaspati:1991nm}. In the present Universe, this strongly interacting phase of matter is believed to exist inside the core of neutron stars, some of which also show strong magnetic fields. The study of such extreme environments in the lab has been pioneered by experiments of \glspl{hic}. These experiments also account for the strongest magnetic fields ever produced on Earth. In peripheral collisions, for instance, the motion of charged particles generates field lines that reinforce each other at the center of the collision. The sizeable impact of such fields on experimentally measurable observables has been recently seen by the STAR Collaboration~\cite{STAR:2023jdd}. In phenomenological studies, theoretical estimates indicate that magnetic fields reach $\sqrt{eB}\sim 0.1\text{ GeV}$ -- for typical Au+Au collisions -- at RHIC energies~\cite{Kharzeev:2007jp} and $\sqrt{eB}\sim 0.5\text{ GeV}$ -- for typical Pb+Pb collisions -- at LHC energies~\cite{Skokov:2009qp}.

Besides strong magnetic fields, high densities can also be created in \glspl{hic}. High-density strongly interacting matter produced in relativistic collisions is relevant not only for ongoing experiments, such as RHIC, but also for future ones, like NICA and FAIR. From a theoretical viewpoint, high densities also provide a path to understanding the different phases of the strong interactions via its underlying microscopic theory: \gls{qcd}.

At vanishing density, the \gls{qcd} phase transition is a smooth crossover~\cite{Aoki:2006we}, whereas at high densities it is conjectured to become first order. The presence of a first-order line would imply the existence of a second-order transition point, known as a \gls{cep}. This hypothesis has been corroborated by analytical approaches, such as the Functional Renormalization Group~\cite{Fu:2019hdw}, Dyson-Schwinger equations~\cite{Fischer:2018sdj}, holographic models~\cite{Hippert:2023bel}, etc.

On the one hand, first-principles numerical simulations of the strong interactions on a lattice -- \gls{lqcd} -- successfully account for all non-perturbative features of \gls{qcd} at vanishing and low densities. On the other hand, the approach becomes unfeasible at moderate and high densities due to the sign (or complex action) problem. Although various techniques have been used to circumvent this issue (see Ref.~\cite{Berger:2019odf} for a recent review), to this day, the determination of the phase structure of \gls{qcd} at high $\mu_B$ from first principles remains a widely open problem.

Theoretically, one can extend the \gls{qcd} phase diagram to the $\mu_B^2<0$ (imaginary) domain, where lattice simulations are sign-problem-free. Interestingly, at imaginary $\mu_B$, the phase diagram has a second-order transition point called Roberge-Weiss \gls{cep}~\cite{Roberge:1986mm}. Also, the \gls{qcd} phase diagram in the $T$-$B$ plane is directly accessible within lattice simulations. In the presence of a magnetic field, the \gls{qcd} crossover becomes sharper and the transition temperature is lowered, suggesting that the crossover would eventually become a first-order transition starting at a \gls{cep}. Moreover, lattice simulations of asymptotically strong magnetic fields found further indication for this scenario~\cite{Endrodi:2015oba}, thus supporting previous expectations based on generic arguments~\cite{Cohen:2013zja} (see also Ref.~\cite{Braguta:2019yci}). Recent \gls{lqcd} results constrained this \gls{cep} to the region: $4 < eB < 9\text{ GeV}^2$, $65<T<95\text{ MeV}$~\cite{DElia:2021yvk}. The impact of magnetic fields on the Roberge-Weiss \gls{cep} has also been studied using \gls{lqcd}, where the findings indicate that this point approaches the $T$-$B$ plane, thus opening the possibility of a connection between the two critical points~\cite{Zambello:2024ucs}.

Another way of probing thermodynamic features of \gls{qcd} is via the \gls{eos}. Moreover, knowing the \gls{eos} as a function of thermodynamic variables ($T$, $\mu_B$, $B$, etc.) is essential to model the hydrodynamic evolution of relativistic collisions across the phase diagram. In this regard, the present knowledge of the interplay between finite densities and magnetic fields is limited. Currently, the \gls{eos} at $B=0$ is known from \gls{lqcd} in a broad range of temperatures at vanishing $\mu_B$ with 2+1 flavors~\cite{Borsanyi:2010cj,HotQCD:2014kol}, as well as at non-vanishing $\mu_B$ with 2+1 flavors~\cite{Bazavov:2017dus}, and 2+1+1 flavors~\cite{Borsanyi:2022qlh}. At $B\neq0$, the \gls{eos} has been calculated with 2+1 flavors at non-zero $\mu_B$ but vanishing strangeness chemical potential~\cite{Astrakhantsev:2024mat}. For a recent review on lattice investigations of the phase diagram and the \gls{eos} at nonzero magnetic fields, see Ref.~\cite{Endrodi:2024cqn}.

\begin{figure}[!h]
\centering
\includegraphics[width=0.6\textwidth]{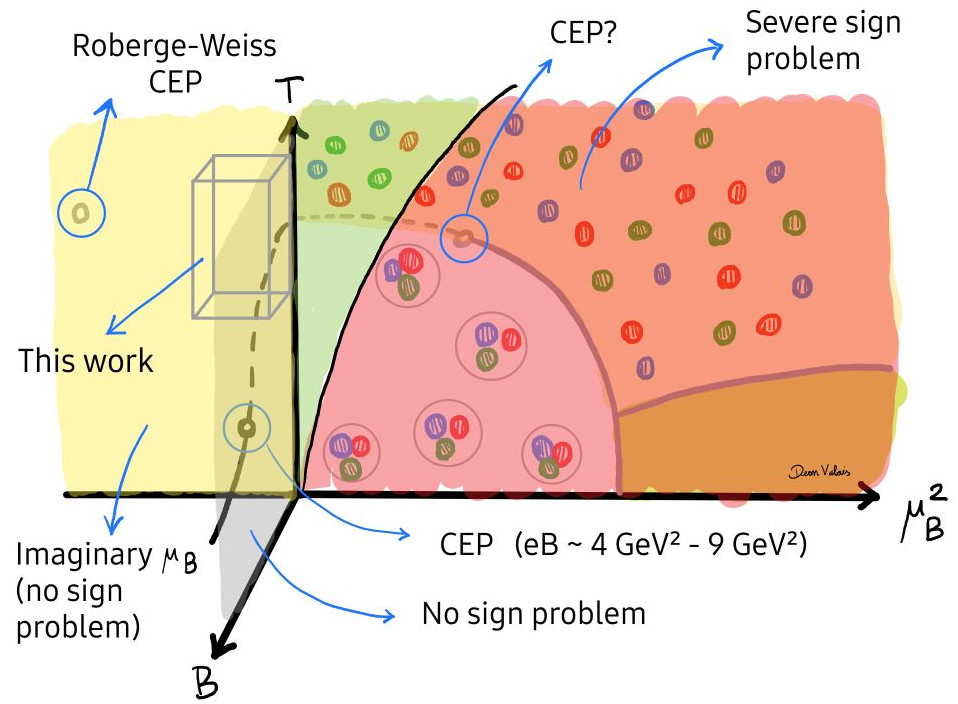}
\caption{Conjectured QCD phase diagram in the $T$-$\mu_B^2$-$B$ space. On the right-hand side, we show the real-$\mu_B$ half-plane, which is challenging due to the sign problem, where a CEP is conjectured to exist. On the left-hand side, the sign-problem-free imaginary-$\mu_B$ half-plane, where the Roberge-Weiss transition point is known. The perpendicular plane depicts the $T$-$B$ phase diagram with a CEP at large values of $B$. The solid indicates the region that we cover with our simulations.}
\label{fig:phase_diagram_qcd}
\end{figure}

Hence, in this proceedings article, we aim to shed light on the \gls{eos} of dense and magnetized \gls{qcd} matter with four dynamical quark flavors. Our method of choice to circumvent the sign problem is analytic continuation from simulations at imaginary $\mu_B$, thus extending our previous work~\cite{Borsanyi:2023buy}, where we carried out simulations at vanishing $\mu_B$. Our \gls{eos} is taken along a trajectory satisfying two constraints motivated by the \gls{hic} setup, namely, strangeness neutrality and isospin asymmetry. Our discussions revolve around two main quantities: the chemical potentials $\mu_Q$ and $\mu_S$ required by our constraints, and the leading-order behavior of the \gls{eos}. We show that the magnetic field significantly impacts the latter, in particular, near the crossover temperature. In Fig.~\ref{fig:phase_diagram_qcd}, we illustrate the conjectured \gls{qcd} phase diagram in the $T$-$\mu_B^2$-$B$ space and indicate where we carried out simulations.

This article is organized as follows:  In Sec.~\ref{sec:simulation_setup}, we describe the simulation setup, including our lattice action and simulation parameters. In Sec.~\ref{sec:eos}, we introduce the definitions of the observables we compute as well as the constraints that we impose on our \gls{eos}. This is followed by Sec.~\ref{sec:constraints}, in which we explain our method to fulfill the constraints at non-zero chemical potential. Finally, in Secs.~\ref{sec:results} and~\ref{sec:conclusions}, we present our results and conclusions, respectively.

\section{Simulation setup}\label{sec:simulation_setup}
In the fermion sector, we employed $2+1+1$ flavors of rooted-staggered fermions with four stout-smearing steps and masses tuned to the physical point. For details on our \gls{lcp}, see Ref.~\cite{Bellwied:2015lba}. In the gauge sector, we used the tree-level Symanzik-improved gauge action~\cite{Weisz:1982zw}. In this setup, we generated gauge configurations on a $32^3\times8$ lattice for various temperatures, imaginary chemical potentials, and magnetic field strengths. We varied the temperature via the inverse gauge coupling $\beta$ according to $T = [a(\beta)N_t]^{-1}$, where $a$ is the lattice spacing and $N_t$ the number of sites in the temporal extent, and the imaginary chemical potential via $\mu_B = i\pi j/8$, with $j\in\mathbb{Z}$. We scanned the temperature in the range $T = 135$--$200$ MeV in steps of $5$ MeV, and the following chemical potentials: $j=0,3,4,5$. Finally, without loss of generality, we considered a uniform magnetic field pointing in the $z$ direction. To simulate the magnetic field, we multiplied the gluon links by the corresponding U(1) factors at each lattice site. For the exact form of these factors, see~\cite{Bali:2011qj}.

In a finite volume, the magnetic field flux satisfies the following quantization condition
\begin{equation}
eB = \frac{6\pi N_b}{L_xL_y}\,,\hspace{0.5cm} N_b\in\mathbb{Z}\,,
\end{equation}
where $L_x$ and $L_y$ are the lattice extents along the $x$ and $y$ directions, respectively. Due to QED-related charge- and magnetic-field renormalizations, the amplitude of the field is typically expressed in terms of the renormalization-group-invariant quantity $eB$. In this fashion, we simulated $eB = 0.3\,,0.5\,,0.8$ GeV$^2$ magnetic fields, which encompass the values expected in the physical systems described in Sec.~\ref{sec:intro}. To achieve the desired values of $eB$ -- given that the change of flux can only be carried out in integer steps $N_b$ -- for each of the aforementioned field strengths, we simulated two values of $N_b$ that surround the desired value. Thus, we obtained results at the desired $B$ using a linear interpolation.

\section{Equation of state and constraints}\label{sec:eos}

Here, we give the definitions of the observables that we compute. Let us start by introducing the pressure in terms of the grand canonical partition function of \gls{qcd} in the rooted-staggered formalism
\begin{equation}
\hat{P} \equiv \frac{P}{T^4} = \frac{1}{VT^3}\ln\Z\,, \hspace{1cm}\Z(\{\hat{\mu}_i\},T,B) = \int\D U e^{-S_g}\prod_f\det[\slashed{D}(\hat{\mu}_f,B)+m_f]^{1/4}\,,
\label{eq:normalized_pressure}
\end{equation}
where $V$ is the spatial volume, $\{\mu_i\} = \mu_u\,,\mu_d\,,\mu_s$, and $\hat{\mu}_f \equiv \mu_f/T$ are the reduced chemical potentials in the quark basis. Notice that we do not include the charm chemical potential although the charm quark is dynamically included in our simulations. Moreover, we define the quark-number susceptibilities as
\begin{align}
\chi^{uds}_{ijk}(\{\hat{\mu}_i\},T,B) = \frac{1}{VT^3}\frac{\partial^{i+j+k}}{\partial\hat{\mu}_u^i\partial\hat{\mu}_d^j\partial\hat{\mu}_s^k}\ln\Z\,. \label{eq: Naive Susc}
\end{align}
These susceptibilities can be converted to the physical basis -- involving baryon, charge, and strangeness quantum numbers -- yielding, for instance, the following number densities
\begin{equation}
n_B = \frac{1}{3}\chi^{uds}_{100} + \frac{1}{3}\chi^{uds}_{010} + \frac{1}{3}\chi^{uds}_{001}\,, \hspace{0.5cm} n_Q = \frac{2}{3}\chi^{uds}_{100} - \frac{1}{3}\chi^{uds}_{010} - \frac{1}{3}\chi^{uds}_{001}\,, \hspace{0.5cm} n_S = -\chi^{uds}_{001}\,.
\label{eq:number_densities}
\end{equation}
Using these quantities, we can now introduce the constraints that we impose on the \gls{eos}, namely,
\begin{align}
n_S=0\,, \hspace{1cm} \frac{n_Q}{n_B}=0.4\,. \label{eq: Isospin asymmetry}
\end{align}
These are known respectively as strangeness neutrality and isospin asymmetry. The ratio $n_Q/n_B$ and the value of $n_S$ are chosen to match the experimental values expected for the byproducts of typical Pb+Pb collisions, for instance. These constraints imply that there is only one independent chemical potential, since $\mu_Q$ and $\mu_S$ can be expressed in terms of $\mu_B$.

Taking these constraints into account, we can compute the pressure change at finite chemical potential along the line of strangeness neutrality and isospin asymmetry using the integral method
\begin{align}
\Delta \hat{P}(\hat{\mu}_B) = \hat{P}(\hat{\mu}_B)-\hat{P}(\hat{\mu}_B=0) = \int_0^{\hat{\mu}_B}\dd\hat{\mu}_B^{\prime}\dv{\hat{P}}{\hat{\mu}_B^{\prime}}\,,
\label{eq:pressure_integral_method}
\end{align}
where $\hat{P}$ was introduced in~\eqref{eq:normalized_pressure}. In terms of the number densities~\eqref{eq:number_densities}, the total derivative of the pressure with respect to $\hat{\mu}_B$ can be written as
\begin{equation}
\dv{\hat{P}}{\hat{\mu}_B} = \qty(n_B + n_Q\pdv{\hat{\mu}_Q}{\hat{\mu}_B} + n_S\pdv{\hat{\mu}_S}{\hat{\mu}_B})\,,
\label{eq:pressure_total_deriv}
\end{equation}
where the partial derivatives of $\hat{\mu}_Q$ and $\hat{\mu}_S$ with respect to $\hat{\mu}_B$ appear due to the constraints. Knowing the right-hand side of Eq.~\eqref{eq:pressure_total_deriv} allows us to compute the pressure shift $\Delta \hat{P}(\hat{\mu}_B)$ via the integral method. Next, we discuss how to tune the charge and strangeness chemical potentials in terms of $\hat{\mu}_B$ such that strangeness neutrality and isospin asymmetry are fulfilled.

\section{Matching the experimental conditions}\label{sec:constraints}
This section focuses on realizing the experimental conditions of \glspl{hic} on the lattice. One way is to precisely tune the simulation parameters. Another way is to compute additional derivatives and extrapolate to strangeness neutrality afterwards. To avoid large computational costs during tuning while still simulating close to the strangeness neutral case to avoid a complicated extrapolation, we combine both. To simulate close to strangeness neutrality at $\hat{\mu}_B(j)$ we estimate $\hat{\mu}_Q(\hat{\mu}_B(j))$ and $\hat{\mu}_S(\hat{\mu}_B(j))$ from all simulations up to $\hat{\mu}_B(j-1)$. 
The shift into exact strangeness neutrality, which is a small difference in $\hat{\mu}_Q$ and $\hat{\mu}_S$. The determination of the shift is performed after the simulation. We determine the new simulation parameters at $\hat{\mu}_B(j)$ by performing extrapolations in $\hat{\mu}_{Q/S}$ using Taylor-expansions up to a certain order, depending on the number of simulation in $\hat{\mu}_B$ direction already performed. This means we make an ansatz such as Eq.~\eqref{eq: Taylor expansion muQ/S}, where $a$ are the parameters we want to determine.
\begin{align}
    \hat{\mu}_{Q/S}(\hat{\mu}_B)= \sum_{i=1}^N  a_{2i+1}\hat{\mu}_B^{2i+1} \label{eq: Taylor expansion muQ/S}
\end{align}
To estimate systematic effects we calculate the parameters $a$ using different approaches. Combining simulations with estimated parameters with a correction to strangeness neutrality afterwards has the advantage that we keep the computational cost for both the simulation and analysis in balance with low statistical uncertainties. The small deviation of the estimated parameters from the true ones in strangeness neutrality ensures that by performing a linear extrapolation we can calculate the shift $\Delta\hat{\mu}_Q$ and $\Delta\hat{\mu}_S$ in the chemical potentials to reach the experimental conditions while keeping the impact on the uncertainties in our observables as small as possible. The susceptibilities in strangeness neutrality after the shift, denoted as $\tilde{\chi}^{BQS}_{ijk}$ read as written in Eq.~\eqref{eq: strangeness neutral susceptibilities}.
\begin{align}
    \tilde{\chi}^{BQS}_{ijk}=\chi^{BQS}_{ijk}+\Delta\hat{\mu}_Q \chi^{BQS}_{i(j+1)k} +\Delta\hat{\mu}_S \chi^{BQS}_{ij(k+1)} \label{eq: strangeness neutral susceptibilities}
\end{align}
We will now discuss the different approaches to estimating the chemical potentials for the simulations
\subsection{Algebraic Determination}\label{subsec: Algebraic determination}
For the algebraic procedure, let us start by taking the derivative of Eq.~\eqref{eq: Taylor expansion muQ/S} with respect to $\hat{\mu}_B$.
\begin{align}
    \frac{d\hat{\mu}_{Q/S(\hat{\mu}_B)}}{d\hat{\mu}_B}=\sum_{i=0}^N (2i+1)a_{2i+1}\hat{\mu}_B^{2i}
    \label{eq: derivative Taylor expansion muQ/S}
\end{align}
Taking $\hat{\mu}_B=0$ we see that the coefficient $a_1$ is given by $\frac{d\hat{\mu}_{Q/S}}{d\hat{\mu}_B}(\hat{\mu}_B=0)$. Since the simulation at $\hat{\mu}_B=0$ is already in strangeness neutrality, we have a given parameter for the Taylor expansion of Eq.~\eqref{eq: Taylor expansion muQ/S}. The expansion to the first order can be used to estimate the $\hat{\mu}_{Q/S}(j=3)$. For the determination of higher order parameters, one can use the measurements of both $\hat{\mu}_{Q/S}$ as well as higher order derivatives of the expansion, in order to solve algebraically for the parameters $a_{2i+1}$. The advantage of this procedure is its computational simplicity and efficiency, which provides a fast result. However, a notable drawback is that higher-order parameters often have large statistical uncertainties, which can significantly impact the results and lead to larger overall errors, particularly when the higher-order parameters are compatible with zero. This effect is particularly pronounced at low temperatures, as shown in Fig.~\ref{fig: NewSimParms}, where obtaining a sufficient amount of reliable statistics is also more computationally expensive.

\subsection{Fit procedure}\label{subsec: fitprocedure}
Another approach would be to perform a global fit and minimize $\chi^2$. For this, we account for the fact that higher-order terms in Eq.~\eqref{eq: Taylor expansion muQ/S} are associated with larger statistical uncertainties. We fit the value of $\hat{\mu}_{Q/S}(\hat{\mu}_B)$ in strangeness neutrality as well as the first derivative $\frac{d\hat{\mu}_{Q/S}(\hat{\mu}_B=0)}{d\hat{\mu}_B}$. The model used is the Taylor expansion~\eqref{eq: Taylor expansion muQ/S}. The expansion order must be adjusted to balance the number of degrees of freedom with the fact that higher-order terms tend to have larger uncertainties, which can make parameters indistinguishable from zero within errors.

\subsubsection{Incorporating the line of constant magnetic field}

As an additional step to further reduce statistical uncertainties, we leverage the fact that the desired values lie on a line of constant magnetic field. We perform a spline interpolation of the values interpolated to the line of constant magnetic field of $\hat{\mu}_{Q/S}$ as well as for their discrete derivatives with respect to changes in $N_b$. Using these interpolated values and their derivatives, we apply a Taylor expansion (linear in $N_b$) to extrapolate the values from the line of constant magnetic field to the nearest integer values of $N_b$ above and below. This is to reduce systematic effects due to the different distances from the two integer values of $N_b$ to the constant magnetic field value for $N_b$. In Fig.~\ref{fig: NewSimParms}  we observe that this approach yields even smaller statistical uncertainties compared to the simple fit results at the integer $N_b$ values.
\subsection{Conclusion of the methods}
\begin{figure}[h]
    \centering
    \includegraphics[width=0.8\textwidth]{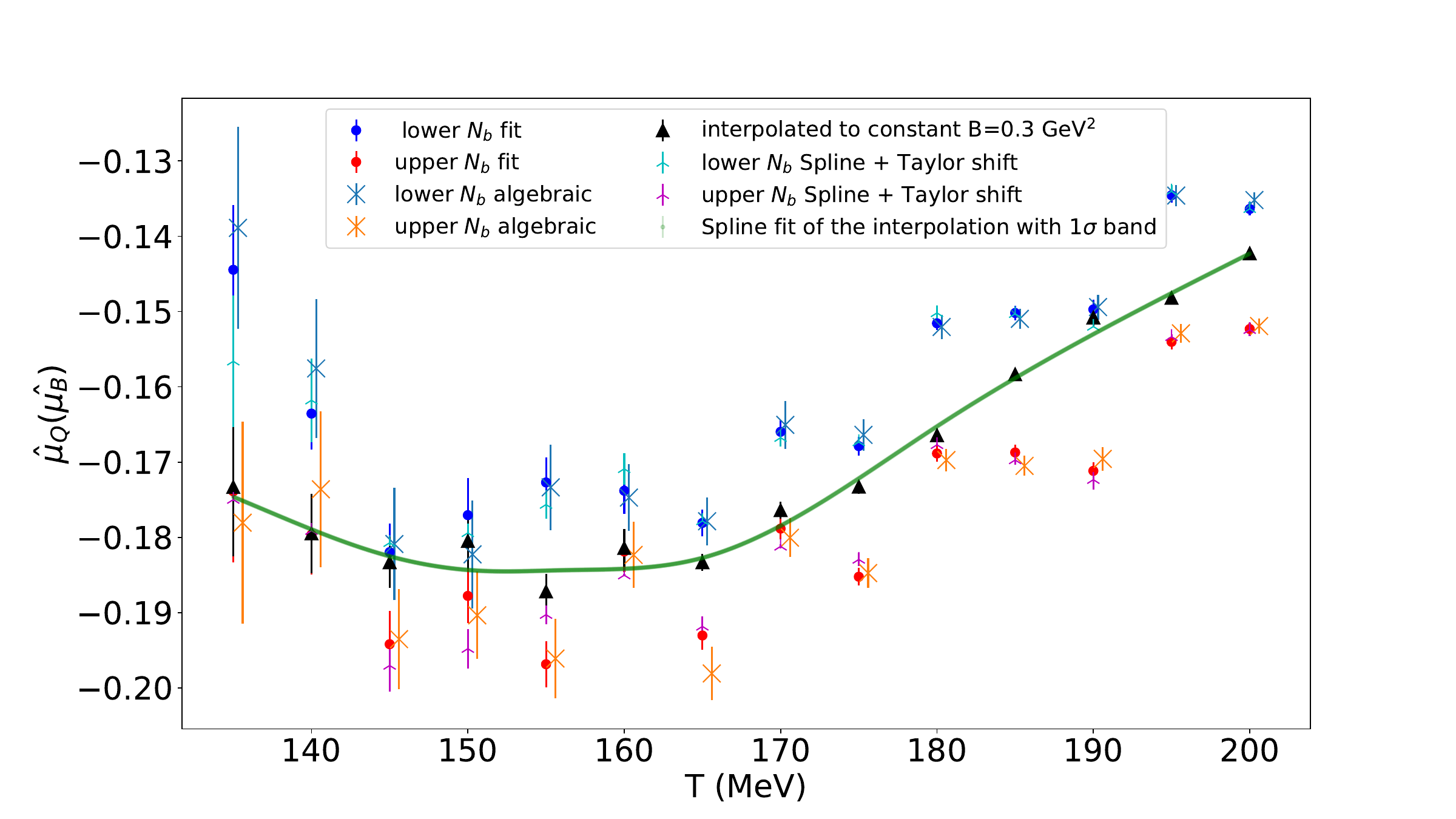}
    \caption{Determined $\hat{\mu}_Q$ for the simulation at $\hat{\mu}_B=\frac{4\pi i}{8}$. The red and blue points represent the determined parameters using the fit method described in (\ref{subsec: fitprocedure}), while the green and yellow crosses follow from the algebraic procedure of subsection (\ref{subsec: Algebraic determination}). The black triangles resemble the interpolated values of the strangeness neutral chemical potentials to the point of constant magnetic field. A spline fit was performed for these values as well as for the discrete derivative from neighboring $N_b$ which was used as input for a Taylor expansion to second order to get the points shown in turquoise and magenta.}
    \label{fig: NewSimParms}
\end{figure}

The results for $\hat{\mu}_Q$ of the described methods are shown in Fig.~\ref{fig: NewSimParms}. The temperature scan with both values for the different $N_b$ at each temperature for the simulation for $\hat{\mu}_B=\frac{4\pi i }{8}$ at an external magnetic field value around $eB=0.3\,$GeV$^2$ is shown. For better visibility, a small shift in the temperature axis was inserted for the different methods. The methods agree within errors for most temperatures. In the low-temperature regime, the overall error is larger for all methods than for the high temperatures. Consequently, the chosen values for the next simulation originate from the spline fit routine, combined with the Taylor expansion to map the results back to integer-valued $N_b$.

\section{Results}\label{sec:results}

\subsection{Chemical potentials}
\begin{figure}[h]
    \centering
    \includegraphics[width=0.8\textwidth]{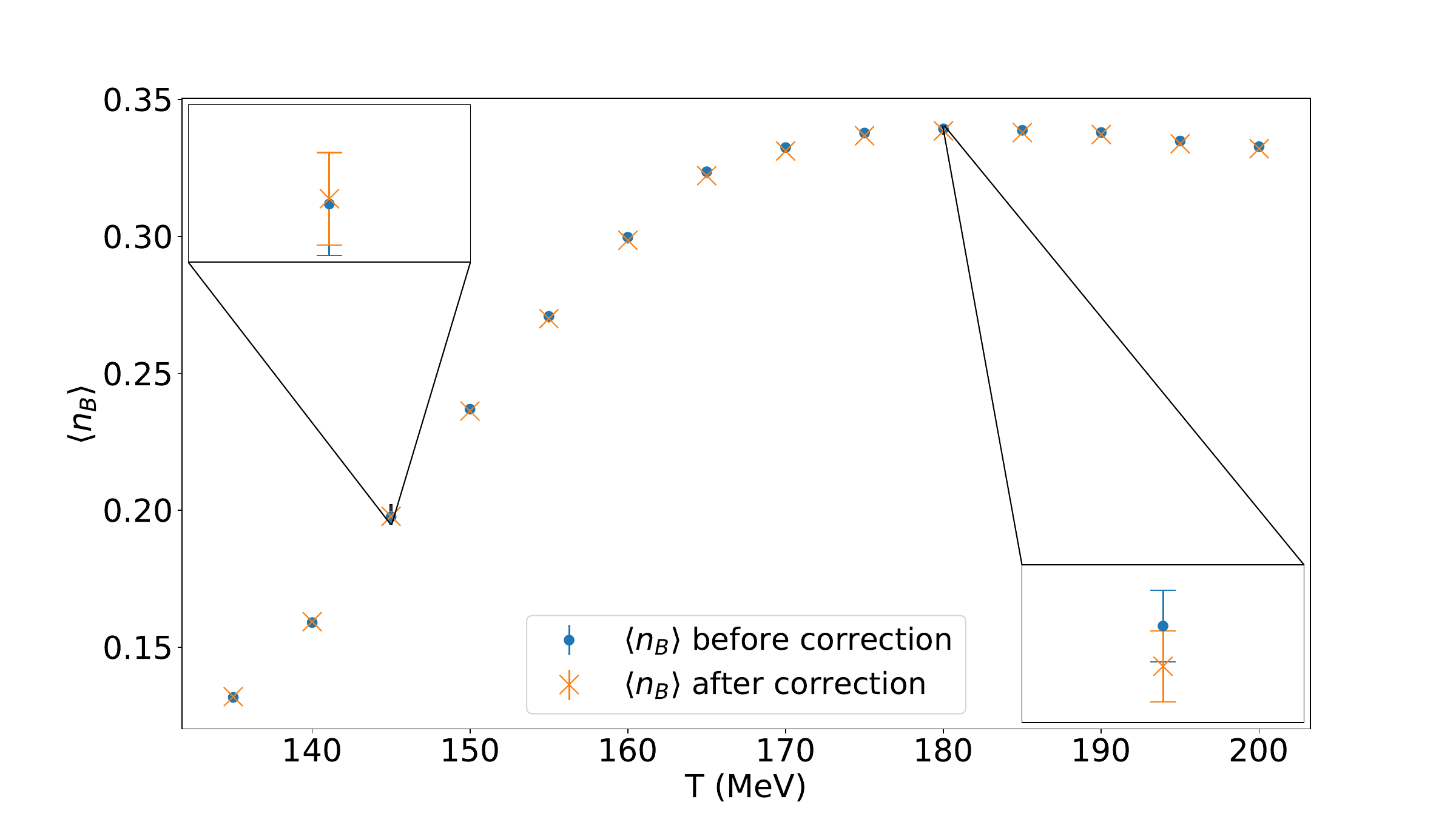}
    \caption{The figure shows the baryon number $\expval{n_B}$ before and after the correction to strangeness neutrality at $eB=0.5\,$GeV$^2$ and $\hat{\mu}_B=\frac{3\pi i}{8}$. At two points we zoom in to show the shifts. We observe that the shift is within errors of the measurement before the correction.}
    \label{fig:baryonnumber}
\end{figure}

In this section, we focus on the impact of the extrapolation to $n_S=0$, as well as the physical interpretation of our findings in the presence of an external magnetic field for simulations with a chemical potential. First, we examine the difference between the baryon number calculated using Eq.~\eqref{eq: Naive Susc}, measured directly after the simulation, and the baryon number after imposing strangeness neutrality and isospin asymmetry calculated via Eq.~\eqref{eq: strangeness neutral susceptibilities}, as shown in Fig.~\ref{fig:baryonnumber}.
\\
The figure confirms two predictions. First of all, we observe that the shift we have to take in order to reach strangeness neutrality is within the errors of the measurements without the shift. Secondly, we observe that the error is manageable -- it does not obscure the trend of the data with respect to temperature -- and remains stable with the shift. This demonstrates that our procedure is sufficient for determining new simulation parameters, where the shift to strangeness neutrality can be effectively handled using a linear Taylor expansion.

\begin{figure}[h]
    \centering
    \begin{subfigure}{0.7\textwidth}
        \includegraphics[width=\textwidth]{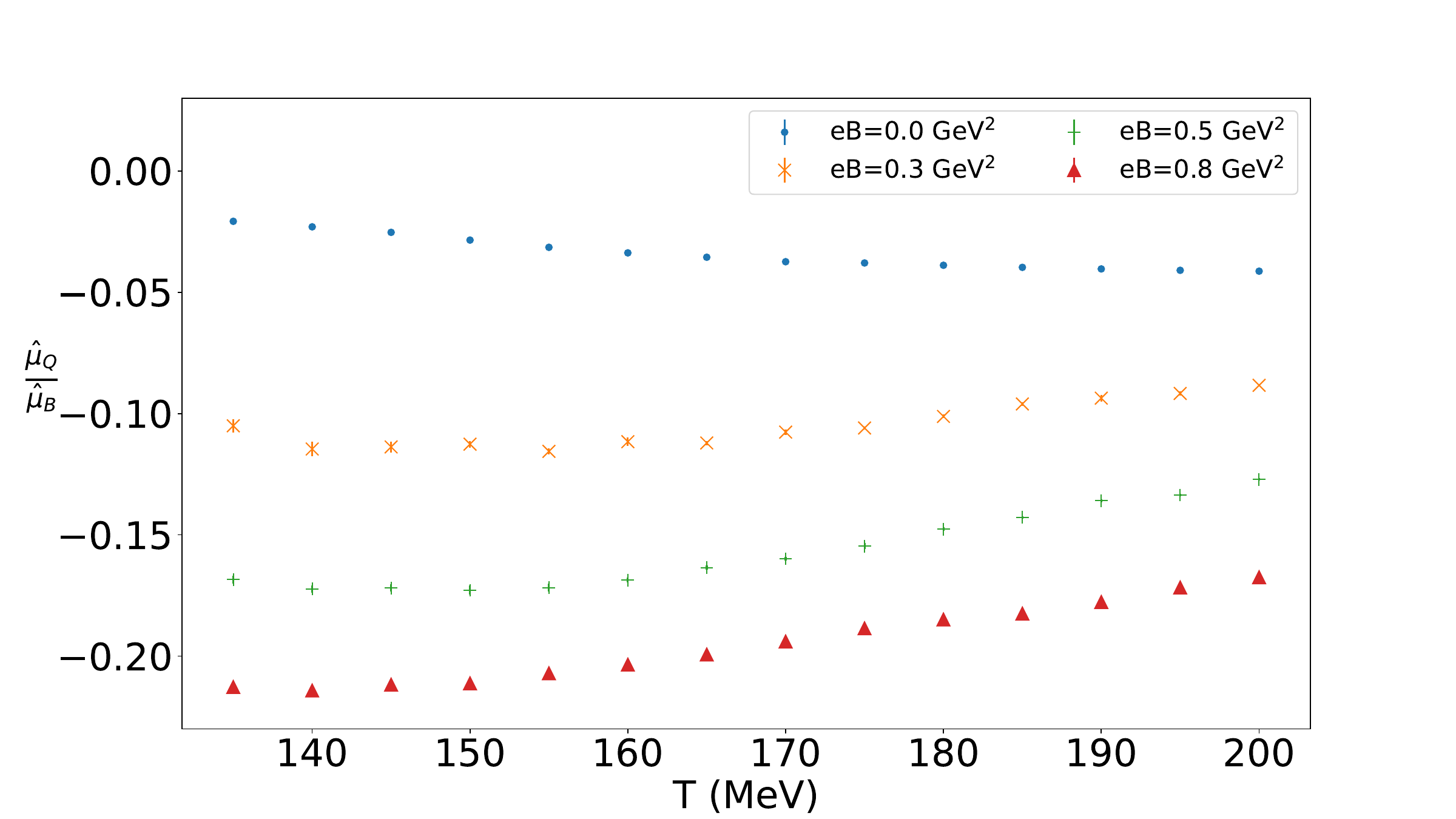}
        \caption{Temperature scan of $\frac{\hat{\mu}_Q}{\hat{\mu}_B}$ for different external magnetic fields.}
        \label{fig:muQ per muB}    
    \end{subfigure}
    \begin{subfigure}{0.7\textwidth}
        \includegraphics[width=\textwidth]{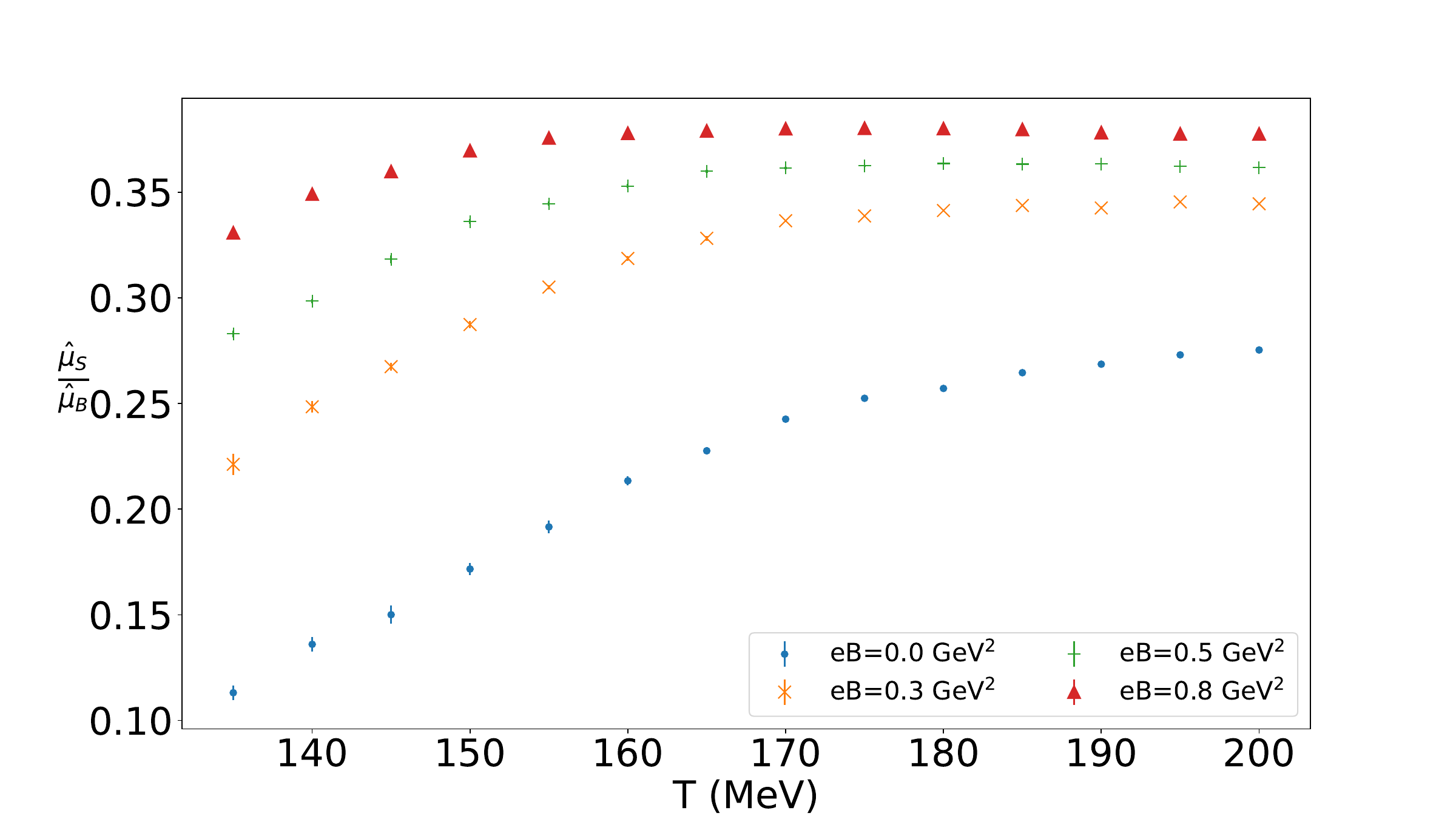}
        \caption{Temperature scan of $\frac{\hat{\mu}_S}{\hat{\mu}_B}$ for different external magnetic fields.}
        \label{fig:muS per muB}
    \end{subfigure}
    \caption{Influence of the external magnetic field on the chemical potentials in the strangeness neutral case. The left figure shows the Temperature scan of the charged chemical potential divided by $\hat{\mu}_B$ at $j=3$ for the different magnetic fields, while the right figure shows the same temperature scan for the different magnetic fields but instead of the charged chemical potential for the strange chemical potential.}
\end{figure}

At the next point, we would like to study how an external magnetic field impacts the chemical potentials, and why. For that let us take a look at the choice of the charge and strange chemical potential per baryon chemical potential for different external magnetic fields in Figs.~\ref{fig:muQ per muB} and~\ref{fig:muS per muB}.
First, let us focus on the Fig.~\ref{fig:muQ per muB}. The overall negative value of $\mu_Q$ originates from the fact that the constraint~\eqref{eq: Isospin asymmetry} for $n_Q$ requires an excess of down quarks over up quarks and, therefore, a suppression of electric charge. 
According to the figure, this negative value decreases consistently throughout the temperature range as the magnetic field grows. To understand this behavior intuitively, let us focus on the low-temperature regime, where the baryon number and electric charge are mostly carried by protons and neutrons. The effect we observe can be attributed to the coupling of these particles to the magnetic field. The lowest energy state is affected by the magnetic field via the spin and the angular momentum terms. Taking into account both of these couplings, together with the proton and neutron gyromagnetic ratios~\cite{CODATA_Recommended_values}, the ground state energies for both hadrons reduce by similar amounts, thus enhancing the abundance of protons and neutrons similarly and pushing the system towards isospin symmetry. To compensate for this, the charge chemical potential needs to be reduced further. This qualitative behavior may be quantified more using the standard hadron resonance gas model approach considered at nonzero magnetic field~\cite{Endrodi:2013cs} and nonzero chemical potentials~\cite{Vovchenko:2024wbg,Marczenko:2024kko}.
A similar argument applies to the strange chemical potential shown in Fig.~\ref{fig:muS per muB}. Here, instead of comparing the gyromagnetic ratios of protons and neutrons, one must examine those of hyperons relative to baryons. This, combined with the requirement of zero net strangeness in the system, leads to a similar conclusion.

\subsection{Analytic continuation}

Finally, in this section, we present preliminary results from our approach to the analytic continuation of the \gls{eos}. Our main observable is shown in Eq.~\eqref{eq:pressure_total_deriv}, i.e.\ the total derivative of the pressure with respect to the baryon chemical potential. The analytic continuation procedure was carried out by fitting a multidimensional spline surface to the quantity $\hat{\mu}_B^{-1}\dd \hat{P}/\dd\hat{\mu}_B$ taking into account the $T$, $B$ and $\hat{\mu}_B^2$ directions. The advantage of this particular combination is that, in leading order, this observable is linear in $\hat{\mu}_B^2$, which permits an extrapolation to the real axis using a linear surface along the $\hat{\mu}_B^2$ direction. Notice that, even though $\dd \hat{P}/\dd\hat{\mu}_B$ vanishes at $\hat{\mu}_B=0$, the quantity $\hat{\mu}_B^{-1}\dd \hat{P}/\dd\hat{\mu}_B$ has a non-zero value there.

\begin{figure}[!h]
    \centering
    \begin{subfigure}{0.49\textwidth}
    \includegraphics[width=\linewidth,trim={0.7cm 0.8cm 0 1.1cm},clip]{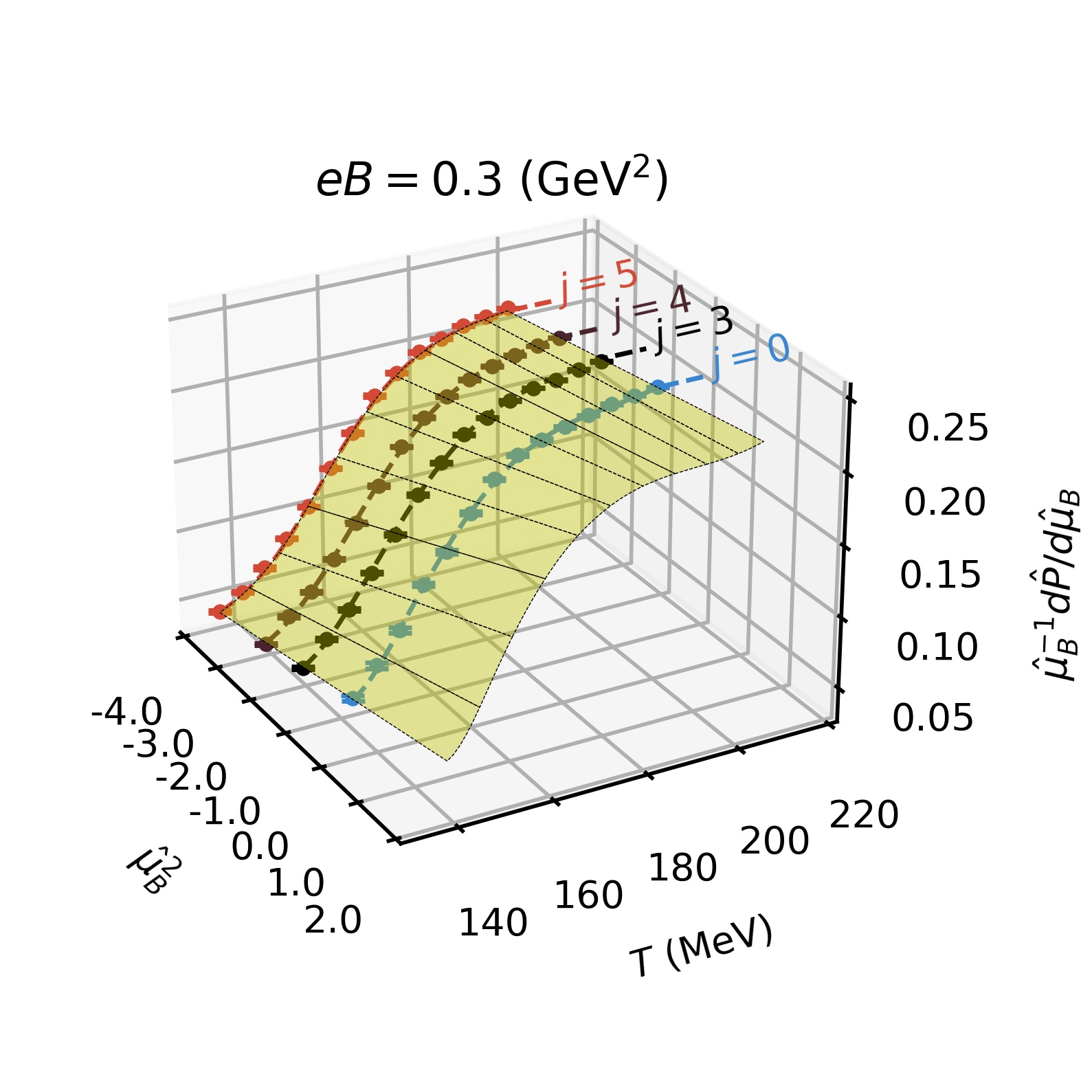}        
    \end{subfigure}
    \begin{subfigure}{0.49\textwidth}
    \includegraphics[width=\linewidth,trim={0.7cm 0.8cm 0 1.1cm},clip]{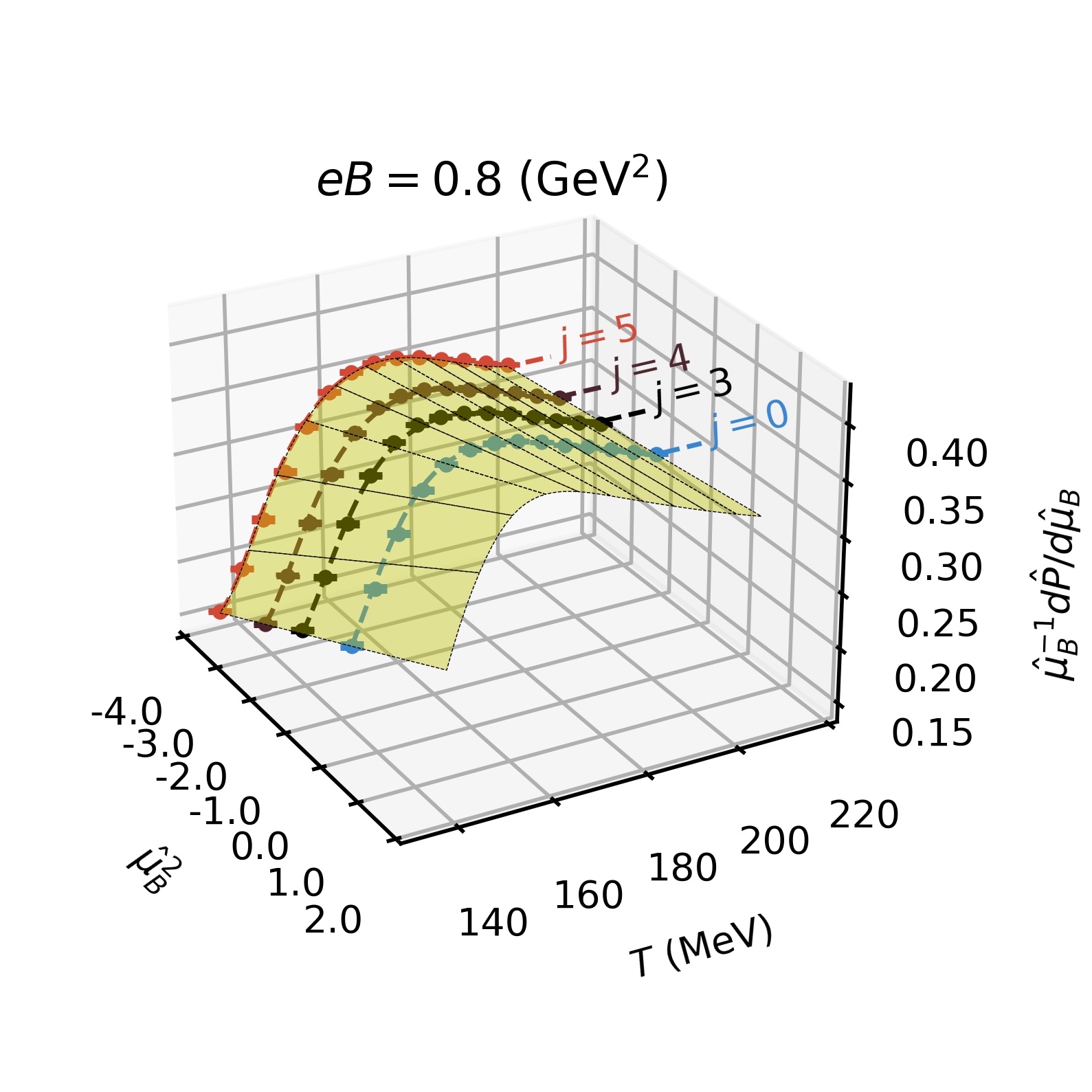}        
    \end{subfigure}
    \caption{Lattice data for $\hat{\mu}_B^{-1}\dd \hat{P}/\dd\hat{\mu}_B$ in the $T$-$\mu_B^2$ plane at $eB = 0.3$ GeV$^2$ (left) and $eB = 0.8$ GeV$^2$ (right). For illustration purposes, we show our preliminary determination of the spline functions (yellow surfaces) that we will use in a future work to carry out the analytic continuation procedure.}
    \label{fig:analytic_continuation_2d}
\end{figure}

The functional form of the spline surface is a fourth-order polynomial in the $T$ and $eB$ directions, whereas it is linear in the $\hat{\mu}_B^2$ direction. This allows us to extract the leading-order behavior of the \gls{eos} in the real-$\hat{\mu}_B$ axis. In order to account for systematic errors in the spline fit, we stochastically generated multiple sets of node points using a Monte Carlo algorithm, for more details see Refs.~\cite{Brandt:2016zdy,Brandt:2022hwy}, where the spline solutions are weighted according to the Akaike criterion~\cite{akaike1100705}. Moreover, to have a more realistic estimation of the systematic effects, we also included spline solutions where the data at $j=5$ were removed from the global fit. In Fig.~\ref{fig:analytic_continuation_2d} we show the result of the multidimensional spline fit for different values of the magnetic field.

The most striking consequence of the non-zero magnetic field, as seen from the surfaces in Fig.~\ref{fig:analytic_continuation_2d}, is the bending towards high temperatures, where the data shows a peak structure near the crossover temperature. Moreover, whereas the behavior of $\hat{\mu}_B^{-1}\dd \hat{P}/\dd\hat{\mu}_B$ is still monotonic at low $B$, it becomes non-monotonic at higher $B$.

\section{Conclusions and outlook}\label{sec:conclusions}

In this proceedings article, we investigated the interplay between finite density and nonzero magnetic fields in \gls{qcd} in the light of observables that can be related to the \gls{eos}. We performed simulations at imaginary chemical potential with 2+1+1 dynamical quarks and a physical pion mass. We used different expansion schemes to constrain our simulation parameters to a strangeness-neutral and isospin-asymmetric line, where we defined $n_Q/n_B = 0.4$ as the isospin asymmetry ratio. These constraints are motivated by phenomenological expectations in \glspl{hic}.

Using a multidimensional spline fit, we took the first steps towards the analytic continuation of our data to the real-chemical-potential axis. At this stage, our results show that the magnetic field has a significant impact on the observables, especially around the crossover temperature. Specifically, we observed that the quantity $\hat{\mu}_B^{-1}\dd \hat{P}/\dd\hat{\mu}_B$ ceased to be monotonic with temperature in the presence of strong enough magnetic fields, namely, for $eB \gtrsim 0.5$ GeV$^2$. This behavior has previously been observed in second-order susceptibilities and in the leading-order coefficient of the \gls{eos}~\cite{Borsanyi:2023buy}. See also Ref.~\cite{Ding:2023bft} for further discussion about the impact of $B$ on these observables.

In order to make contact with continuum \gls{qcd} physics, the continuum limit of the results presented here must still be carried out, as well as the full analytic continuation of our data. Nevertheless, our current findings reveal interesting features of the impact of strong magnetic fields on hot and dense \gls{qcd} matter which might be insightful for the modeling of systems where such medium is created: \glspl{hic}.\\

\acknowledgments

This work is supported by the
MKW NRW under the funding code NW21-024-A. The authors gratefully acknowledge the Gauss Centre for Supercomputing e.V. for funding this project by providing computing time on the GCS Supercomputer
SuperMUC-NG at Leibniz Supercomputing Centre. The authors are also grateful to Gergely Mark\'{o} for discussions that shaped our approach to the results presented here. Moreover, we acknowledge Arpith Kumar and Jin-Biao Gu for helpful discussions on dense and magnetized \gls{qcd} during this conference.
Gergely Endr\H{o}di acknowledges funding by the Hungarian National Research, Development and Innovation Office (Research Grant Hungary 150241) and the European Research Council (Consolidator Grant 101125637 CoStaMM).
\begin{multicols}{2}
\printglossary[type=\acronymtype,nonumberlist]
\end{multicols}

\bibliographystyle{utphys}
\bibliography{bibliography.bib}
\end{document}